\documentclass[twocolumn, graphics, floatfix, a4paper, aps, prx, superscriptaddress, longbibliography, showpacs,citeautoscript]{revtex4-2}
\usepackage{graphicx}
\usepackage{bm}
\usepackage[dvipsnames]{xcolor}
\usepackage{amssymb,amsmath}
\usepackage[normalem]{ulem}
\usepackage{booktabs}
\usepackage{longtable}
\usepackage{braket}
\usepackage[T1]{fontenc}
\usepackage[utf8]{inputenc}
\usepackage[english]{babel}
\usepackage{xspace}
\usepackage{epstopdf}
\usepackage[final]{microtype}
\usepackage{ragged2e}
\usepackage[colorlinks=true,urlcolor=blue,citecolor=blue,linkcolor=blue]{hyperref}

\def\pt{\partial}

\def\rr{\mathbf{r}}
\def\qq{\mathbf{q}}
\def\aa{\mathbf{a}}
\def\bb{\mathbf{b}}

\def\PH{\mathcal{P}}
\def\bsigma{\boldsymbol{\sigma}}

\def\TRS{T}

\def\MR{moir\'e}

\newcommand{\qql}{\textquotedblleft}
\newcommand{\qqr}{\textquotedblright\xspace}
\newcommand{\vc}[1]{\bm{\mathrm{#1}}}

\newcommand{\dbk}[2]{\left\langle #1 | #2 \right\rangle}

\graphicspath{{./Figures/}}

\begin{document}
\title{Superfluidity and Quantum Geometry in Twisted Multilayer Systems}
\author{P\"aivi T\"orm\"a}
\email{paivi.torma@aalto.fi}
\affiliation{Department of Applied Physics, Aalto University School of Science,
FI-00076 Aalto, Finland}

\author{Sebastiano Peotta} 
\affiliation{Department of Applied Physics, Aalto University School of Science, FI-00076 Aalto, Finland}

\author{Bogdan A. Bernevig}
\affiliation{Department of Physics, Princeton University, USA}
\affiliation{Donostia International Physics Center, P. Manuel de Lardizabal 4, 20018 Donostia-San Sebastian, Spain}
\affiliation{IKERBASQUE, Basque Foundation for Science, Bilbao, Spain}

\begin{abstract}
Designer 2D materials where the constituent layers are not aligned may result in band structures with dispersionless, \qql flat\qqr bands. Twisted bilayer graphene has been found to show correlated phases as well as superconductivity related to such flat bands. In parallel, theory work has discovered that superconductivity and superfluidity of flat band systems can be made possible by the quantum geometry and topology of the band structure. These recent key developments are merging to a flourishing research topic: understanding the possible connection and ramifications of quantum geometry on the induced superconductivity and superfluidity in moir\'e multilayer and other flat band systems. This article presents an introduction to how quantum geometry governs superfluidity in  platforms  including,  and beyond, graphene. We explain how a new type of  topology discovered in twisted bilayer graphene could affect superconductivity, pinpoint the geometric contribution in its Berezinskii-Kosterlitz-Thouless critical temperature, and mention \MR{} materials beyond twisted bilayer graphene. Ultracold gases are introduced as a complementary platform for quantum geometric effects and a comparison is made to moir\'e materials. An outlook sketches the prospects of twisted multilayer systems in providing the route to room temperature superconductivity.

\end{abstract}

\maketitle

\section{Introduction}
\label{sec:introduction}

Superconductivity or superfluidity require bosonic particles to be in the same energy state; in case of superconductors these bosons are composites of two electrons, so-called Cooper pairs. However, being in the same state is not alone sufficient. The system needs to be able to host stable supercurrents, that is, dissipationless flow of particles: this is what makes superconductors useful in generating large magnetic fields or in providing qubits for quantum computing. It has long been the holy grail of condensed matter physics to bring the phenomenon of superconductivity to room temperature, and research on high-$T_{\rm c}$ materials has been leading the way~\cite{Zhou2021HighTc}. However, a complementary route is now emerging, that of purposefully designing materials less complex, and more tunable, than the traditional high-$T_{\rm c}$ ones, following simple but fundamental theory guidelines.    

The first theory suggestion for achieving high temperature superconductivity is to use bands where the energy dispersion as function of momentum, $\varepsilon(\vc{k})$, is constant -- so-called flat bands. The critical temperature $T_{\rm c}$, at which Cooper pairs form, is predicted~\cite{kopnin2011,heikkila2011,Khodel1994}
to be linearly proportional to the interaction $U$ between the Cooper pair constituents. Comparison to the well-known prediction of the Bardeen-Cooper-Schrieffer (BCS) theory of superconductivity~\cite{Schrieffer1964}, $T_c \propto e^{-{1/(n_F U)}}$ ($n_F$ is the density of states at the Fermi level) shows that the critical temperature is \textit{exponentially enhanced} compared to dispersive systems, in the BCS formalism. The enhancement is due to high density of states and dominance of interactions over kinetic energy.  The band does not need to be exactly flat to benefit from this; any band where the interaction $U$ is much larger than the bandwidth will do. 

A second guideline from theory, however, is needed since the flat band provides only the critical temperature for Cooper-pairing but does not determine whether superfluidity and supercurrents exist. Their existence is not obvious in a flat band: if non-interacting fermionic particles are localized on atomic sites and their ground state is an insulator, then no current is possible. Would interacting particles as well just form localized pairs which cannot move? Or, is supercurrent possible, and if yes, under which general conditions? The answer was found only relatively recently: stable supercurrent and superfluidity are possible even in a flat band, if the band has non-trivial \textit{quantum geometry}~\cite{peotta2015,julku2016,liang2017,torma2018,Huhtinen2021two}. The quantum geometry here refers to distances and curvatures in the space (or manifold) formed by the electronic Bloch functions, i.e. the eigenfunctions of the band.

The quantum geometry of a band is characterized by the quantum geometric tensor (QGT)~\cite{Provost:1980,resta2011}. Its real part is the quantum metric, which quantifies the amplitude distance between two close quantum states. Quantum states have, however, also a phase difference, leading to the definition of the Berry phase. Indeed, the imaginary part of the QGT is the Berry curvature. Integrated over the Brillouin zone, the Berry curvature gives the Chern number -- an integer -- a topological invariant that changes only stepwise. Non-zero values of the quantum metric or of the Berry curvature make the quantum geometry of the system non-trivial. If the Chern number is non-zero, the system is also topologically non-trivial. These concepts are defined in Box 1. The Berry curvature is known to be relevant for numerous physical phenomena, including topological insulators and superconductors~\cite{Hasan2010,bernevig_topological_2013}, while the importance of the quantum metric is only emerging -- it turns out to be the quantity that governs flat band superconductivity. 

\begin{figure*}[h]
\begin{minipage}[s]{\textwidth}
\justifying
\includegraphics[width=0.8\columnwidth]{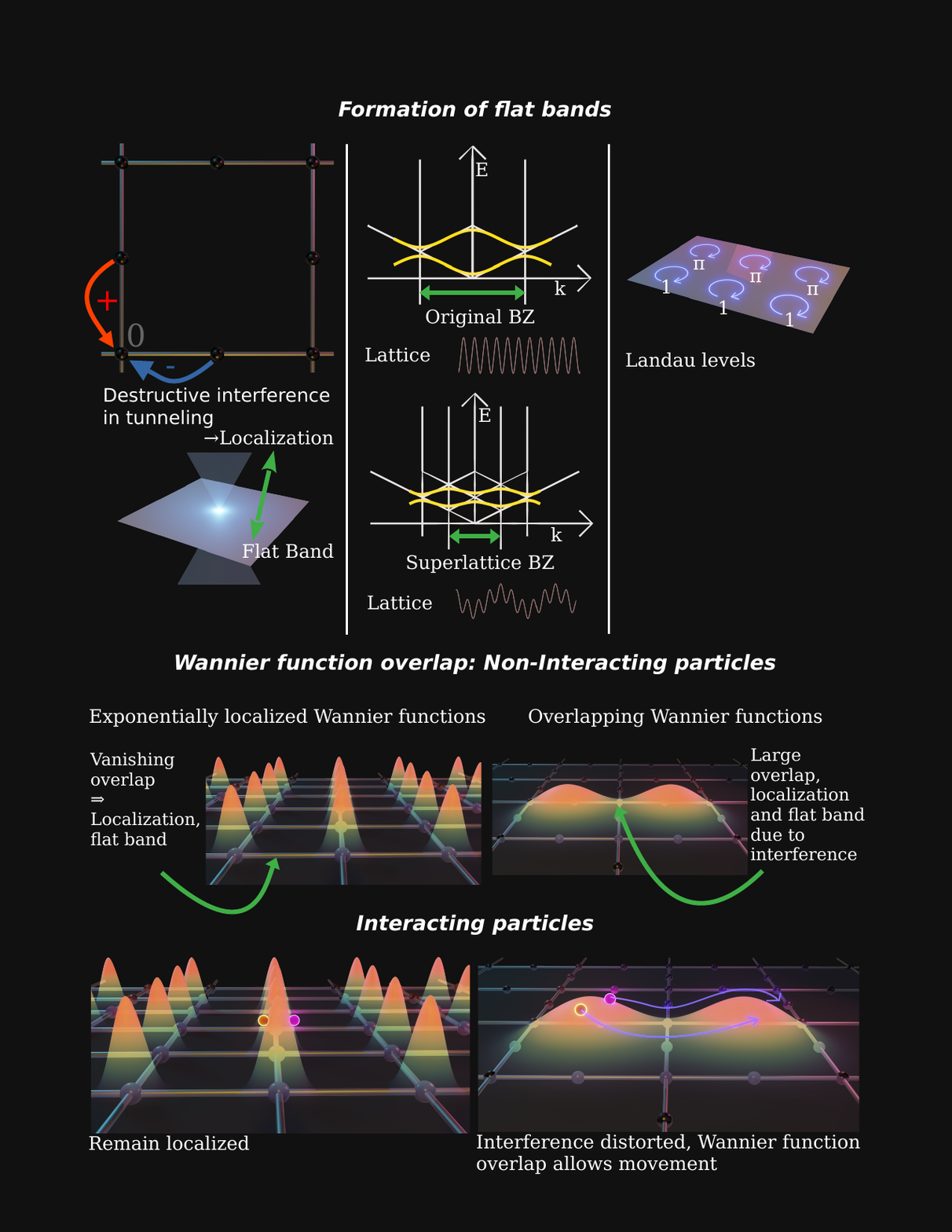}
\caption{\label{fig:figure1}\textbf{Formation of flat bands and the role of Wannier function overlap in flat band superfluidity.} Upper panel: (left) Certain lattice geometries, such as the Lieb lattice example, support eigenstates that occupy a subset of lattice sites with such phases of the wavefunction that tunneling to the rest of the sites is impossible due to destructive interference, leading to localized states and a flat band. (middle) Introducing a superlattice on top of a periodic system, a BZ smaller than the original emerges, and if new band gaps are opened, band flattening takes place. (right) Landau levels are a well-known example of a flat band system. One can understand the localization and band flattening via plaquettes with winding phases so that interference prevents movement, or alternatively, one may think the plaquettes as forming a larger unit cell and a superlattice. Lower panels: Non-interacting particles can be immobile either if their Wannier functions are exponentially localized at the lattice sites, or due to destructive interference. In the former case, also interacting particles will be localized, while in the latter, interactions distort the interference and the particles can move due to the finite overlap of the Wannier functions. Wannier function overlaps are controlled by the quantum geometry of the band; in particular, topological systems cannot have exponentially localized Wannier functions. Picture by Antti Paraoanu.}
\end{minipage}

\end{figure*}

The underlying reason why quantum geometry is important for superconductivity and superfluidity is that it quantifies the \textit{overlaps between the Wannier functions} of a band (Section~\ref{Wfunctionoverlap}). If the Wannier functions are very localized and do not overlap, the flat band is trivial and transport is not possible. Indeed, a trivial flat band is equivalent to an insulator in the atomic limit of large atom-atom separation. However, there are many systems where flat bands occur in conjunction with mobile -- not atomic --  electrons. Many lattice geometries, such as -- but not limited to -- the Lieb lattice, support eigenstates with opposite phases at nearby lattice sites, so that tunneling contributions from neighboring sites interfere destructively. This leads to a set of degenerate eigenstates, i.e.~a flat band. Another route to nearly flat bands are superlattices that introduce new, smaller Brillouin zones (BZ), and consequently new band gaps that then lead to band flattening; see Fig.~\ref{fig:figure1} upper panel. Band gap opening is given, in general, by the interference between a state of certain momentum and a backscattered state with opposite momentum. Therefore both mechanisms of flat band generation rely on the quantum interference of the wave functions. One is best illustrated in the real space and the other in the momentum space. Many types of lattices (line-graph lattices, split graph, Lieb, etc.~\cite{mielke1991a, lieb1989,Leykam2018}) can support flat bands which result from wavefunction interference. A large number of such-obtained flat bands are  topological in origin~\cite{calugaru2021general} and exhibit nonzero (non-Abelian) Berry curvature. Whether or not adding interaction to these bands leads to transport and supercurrent depends on the overlap of the Wannier functions, that is, the Fourier transforms of the Bloch functions, see Fig.~\ref{fig:figure1} lower panel. The overlap, in turn, is governed by quantum geometry. This explains the quantum geometric origin of flat band superconductivity. 

In the quest for room temperature superconductivity, one thus needs to design not only the band dispersion but also the properties of the Bloch functions, that is, the band geometry and topology. Twisted bilayer graphene (TBG) and other moir\'e materials, where a superlattice is created by a twist angle between two graphene layers, offer excellent potential for flat band engineering and control~\cite{MacDonald2019,Andrei2020,Balents2020,Kennes2021,Andrei2021}. At about one degree, so-called first magic angle, a set of flat bands appears in the electronic spectrum. The individual graphene sheets have finite Berry curvatures near the four Dirac points of TBG (two per layer, located at the $K$ and $K'$ high-symmetry points of the BZ called valleys). Inter-layer hopping of electrons \emph{couples} the Dirac points of the \emph{same} valley and Berry curvature (also called here "helicity") in the two layers. In contrast, the \emph{different} valleys are \emph{decoupled}. The coupling of the same valley, same helicity Dirac nodes, with their additive helicity, leads to the emergence of a new type of topology of the single-valley model of TBG. Further degrees of freedom in the design of bands and their topology are given by stacking more than two graphene layers, or creating moir\'e materials composed of other elements than carbon. Quantum gases of ultracold atoms form another playground for the study of interacting flat bands. In these systems, one can create moir\'e geometries as well as simpler flat band systems such as Lieb and kagome lattices, and tune the interparticle interaction over a wide range from repulsive to attractive. 

In this review, we first introduce the essential theory of quantum geometry-dependence of superfluidity, and show how a finite Chern number of a band gives a fundamental lower bound of the flat band superfluid weight, the observable that quantifies the ability of the system to support superfluid transport. We then explain how a similar lower bound is given by a new type of zero-Chern-number topology in the context of TBG, and review theory work that predicts the relevance of quantum geometric effects in the experimentally observed TBG superconductivity and other materials. We then review the work on moir\'e systems in ultracold gases. We discuss the prospects of reaching room temperature superconductivity by band structure and topology design, and briefly refer to the exciting possibilities available for bosonic condensates of atoms, polaritons or photons in flat bands. 

\begin{figure*}[h]
\begin{minipage}[s]{\textwidth}
\justifying
\textbf{Box 1: Band structure invariants and topology}

The eigenfunctions of a Bloch band, describing electrons in crystalline materials, take the generic form $\ket{\psi_{n\vc{k}}} = e^{i\vc{k}\cdot\vc{r}}\ket{u_{n\vc{k}}}$, where $n$ denotes the band number and $\ket{u_{n\vc{k}}}$ is  the periodic Bloch function since it has the same periodicity as the crystalline lattice. They are defined up to a $\vc{k}$-dependent phase, and can enter the expression of an observable quantity only through band structure invariants, which do not depend on how this phase is chosen (called a gauge choice).

A basic invariant is the \textbf{\textit{quantum geometric tensor (QGT)}}
\begin{equation}\label{abelianQGT}
\begin{split}
\mathcal{B}_{ij}(\vc{k}) & = 2\mathrm{Tr}\big[P(\vc{k})\partial_{i}P(\vc{k})\partial_jP(\vc{k})\big]\,.
\end{split}
\end{equation}
It is expressed in terms of the projector to the band $n$ of interest, $P(\vc{k}) = \ket{u_{n\vc{k}}}\bra{u_{n\vc{k}}}$, which is unchanged under the gauge transformation $\ket{u_{n\vc{k}}} \to e^{i\theta(\vc{k})}\ket{u_{n\vc{k}}}$.
Importantly, the QGT  is a Hermitian matrix with non-negative eigenvalues, i.e.~a positive semi-definite complex matrix. A positive semi-definite complex matrix $A_{ij}$ has the property that $\sum_{ij}b_i^*A_{ij}b_j \geq 0$ for arbitrary complex numbers $b_i$. 

Using $P(\vc{k}) = P^\dagger(\vc{k}) = P^2(\vc{k})$, the real part of the QGT can be written as
\begin{equation}
g_{ij}(\vc{k}) = \mathrm{Re}\,\mathcal{B}_{ij}(\vc{k}) = \mathrm{Tr}[\partial_iP(\vc{k})\partial_jP(\vc{k})]   
\end{equation}
and is known as the \textbf{\textit{quantum metric}} (or Fubini-Study metric). It is a measure of the distance between infinitesimally close  wave functions in $\vc{k}$-space.  The quantum metric appears also as the gauge-invariant part of the Marzari-Vanderbilt localization functional for Wannier functions (see Sec.~\ref{Wfunctionoverlap}).

The imaginary part of the QGT is related to the concept of Berry (geometric) phase. Let us first remind about the Berry connection in band $n$
\begin{equation}
 \vc{A}(\vc{k}) = i \braket{u_{n\vc{k}} | \bm{\nabla}_{\vc{k}}u_{n\vc{k}} } \,. 
\end{equation}
The Berry phase can be expressed as an integral over a surface $S$ whose boundary is a closed curve $\gamma = \partial S$
\begin{equation}
\Phi_{\rm Berry} = \int_{S} d\vc{S}\cdot \bm{\nabla}_{\vc{k}} \times \vc{A}(\vc{k}) = \frac{1}{2}\int_{S} dS_l\varepsilon^{lmn}\mathrm{Im}\,\mathcal{B}_{nm}(\vc{k})\,.
\end{equation}
The quantity $\bm{\nabla}_{\vc{k}}\times \vc{A}(\vc{k})$, proportional to the imaginary part of the QGT,
is known as the \textbf{\textit{Berry curvature}}. If in the above equation the integral is extended to the whole Brillouin zone (assuming dimension $d = 2$) one obtains the so-called \textbf{\textit{Chern number}}, a topological invariant,
\begin{equation}
\mathcal{C} = \frac{1}{2\pi} \int_{\rm BZ}d^2\vc{k}\, \mathrm{Im}\, \mathcal{B}_{12}(\vc{k})\,.
\end{equation}
The Chern number can take only integer values. 

The fact that the QGT is positive semi-definite implies that the real and imaginary parts satisfy a number of constrains, of which Eq.~\eqref{eq:ineq} is an example. The Berry connection $\vc{A}(\vc{k})$ and curvature can be generalized to the case of multiple bands ($\vc{A}^{mn}(\vc{k}) = i \braket{u_{m\vc{k}} | \bm{\nabla}_{\vc{k}}u_{n\vc{k}} }$), in which case they are called non-Abelian and bring about a wide array of new topological invariants. One example is the \textbf{\textit{Euler class}}, $e_2$, a topological invariant expressed in terms of the non-Abelian Berry curvature. In systems with $C_{2z}\TRS$ symmetry, the non-Abelian Berry connection and curvature of two bands can  be written as $\mathbf{A}(\mathbf{k}) = -\mathbf{a}(\mathbf{k})\sigma_y$ and $\mathcal{F}_{xy}(\mathbf{k}) = -f_{xy}(\mathbf{k})\sigma_y$, respectively ($\sigma_y$ is a Pauli matrix).  The Euler class $e_2$  is the invariant defined as 
\begin{equation}
\label{eq:euler_class}
	e_2 = \frac{1}{2\pi}\int d^2k\,f_{xy}\,.
\end{equation}
 The abelian QGT in Eq.~\eqref{abelianQGT} is the trace of a more general,  positive-definite \textbf{\textit{non-Abelian QGT}} (matrix) $\mathfrak{G}_{ij}$: if the Bloch states for $n=1\ldots N$ bands are written as a vector $u({\bf{k}}) =(|u_{1\bf{k}}\rangle,\ldots, |u_{N\bf{k}}\rangle)$, the non-Abelian QGT is $\mathfrak{G}_{ij} = \partial_i u^\dagger (1- u u^\dagger ) \partial_j u$ and $\mathcal{B}_{ij} =Tr[\mathfrak{G}_{ij}]$. 
\end{minipage}

\label{fig:box1}
\end{figure*}

\section{The quantum geometric contribution of superfluid weight: topological lower bound for supercurrent}
\label{sec:quantum_geometry}

\subsection{Superfluid weight}

The electrodynamic properties of superconducting materials are captured by the constitutive relation (in London gauge)~\cite{Schrieffer1964,Tinkham2004} 
	\begin{equation}
		\label{eq:London_equation}
		\vc{j} = -D_{\rm s}\vc{A}\,,
	\end{equation}
	with $\vc{j}$ the current density and $\vc{A}$ the vector potential.
	In tandem with Maxwell's equations, Eq.~\eqref{eq:London_equation} provides a quantitative description of the phenomena of perfect conductivity and perfect diamagnetism (Meissner effect), therefore a nonzero superfluid weight $D_{\rm s}\ne 0$ is the very criterion of superconductivity~\cite{Scalapino1992,Scalapino1993}.
	The superfluid weight measures also the energy required to create a modulation of the phase $\phi(\vc{r})$ of the superconducting order parameter $\psi(\vc{r}) = |\psi(\vc{r})|e^{i\phi(\vc{r})}$ since the free energy contains the term
	\begin{equation}
		\label{eq:free_energy_Ds}
		\Delta F = \frac{\hbar^2}{2e^2}\int d^d{\vc{r}}\, \sum_{ij = x,y,z} D_{{\rm s},ij}\partial_{i}\phi(\vc{r}) \partial_{j}\phi(\vc{r})\,, 
	\end{equation}
	with $d = 2,3$ the spatial dimension, $\hbar$ the Planck's constant and $e$ the electron charge.
	In anisotropic systems the superfluid weight is not a scalar but a rank-2 tensor $D_{{\rm s},ij}$ and Eq.~\eqref{eq:London_equation} is modified accordingly.
	
 	The superfluid weight of a spinful electron band at zero temperature given by single-band BCS theory is~\cite{Schrieffer1964,Chandrasekhar1993}
	\begin{equation}
		\label{eq:Ds_BCS_single_band}
		\begin{split}
			&D_{{\rm s},ij} = \frac{e^2}{\hbar^2} \int \frac{d^d\vc{k}}{(2\pi)^d} \, f\big(\varepsilon(\vc{k})\big)\frac{\partial^2 \varepsilon(\vc{k})}{\partial k_i \partial k_j}\,,
		\end{split}
	\end{equation}
	where $\varepsilon(\vc{k})$ is the dispersion of the band.   Here and in the following, quasimomentum $\vc{k}$ integrals are performed over the first BZ.
	The function $f(\varepsilon)$ is the occupation of a single-particle state with energy $\varepsilon$ in the BCS ground state~\cite{Schrieffer1964,Tinkham2004}.
	By assuming an approximately parabolic band $\varepsilon({\vc{k}}) \approx \hbar^2\vc{k}^2/(2m_{\rm eff})$, one obtains $D_{\rm s}  = e^2n/m_{\rm eff}$, where $n$ is the total number density and $m_{\rm eff}$ the particle effective mass~\cite{Leggett1998}.
	At finite temperature only a fraction of the charge carriers participate to superfluid transport, thus the superfluid weight reads $D_{\rm s} = e^2n_{\rm s}/m_{\rm eff}$ where $n_{\rm s}$ is the superfluid density and $0 \leq n_{\rm s}/n \leq 1$ the corresponding superfluid fraction~\cite{London1935,Tinkham2004,Schrieffer1964}. In the extreme limit in which the single-particle effective mass is very large or even diverging, the superfluid weight should be very small or even vanishing according to Eq.~\eqref{eq:Ds_BCS_single_band}~\cite{Basov2011}. To best understand why this is not generally the case in multiband systems, it is illuminating to first consider the two-body problem in a flat band~\cite{torma2018}.
	
	\subsection{Geometric origin of the pair effective mass in a flat band}
	\label{section:pairmass}
	
	The Fermi sea is unstable towards pair formation under an arbitrarily small attractive interaction between electrons~\cite{Schrieffer1964}, even when in three dimensions the strength of a short-range attractive interaction needs to be larger than a threshold for two quantum particles to form a bound state. On the contrary, in the limit of infinite effective mass, that is in a flat band, a bound state is always present for any small value of the attractive interaction strength~\cite{torma2018}.
	A crucial question is: what is the effective mass of the bound state? Naively, one could expect that since the constituent particles have infinite effective mass this would be the case also for the bound state they form. This would mean that the superfluid weight is zero and no superconducting state can occur in a partially filled flat band. However, one can show that the bound state can disperse as a result of interactions~\cite{torma2018}.
	
To illustrate this important point we consider a generic lattice model with Hamiltonian $\hat{H} = \hat{H}_0 + \hat{H}_{\rm int}$ and introduce some definitions. The  term $\hat{H}_{0} = \sum_{\vc{i}\alpha,\vc{j}\beta} \sum_\sigma \hat{c}_{\vc{i}\alpha\sigma}^\dagger K^\sigma_{\vc{i}\alpha,\vc{j}\beta} \hat{c}_{\vc{j}\beta\sigma}$ describes the single-particle hopping between the lattice sites. The  hopping amplitudes $K^\sigma_{\vc{i}\alpha,\vc{j}\beta}$ and the fermionic field operators $\hat{c}_{\vc{i}\alpha\sigma}$ are labelled by the unit cell index $\vc{i} = (i_1,i_2,i_3)$, and the sublattice (orbital) index $\alpha = 1,\dots,N_{\rm orb}$ ($N_{\rm orb}$ is the number of orbitals in a unit cell), while $\sigma =\uparrow,\downarrow$ is the spin index. For concreteness we focus on an interaction term of the Hubbard form $\hat{H}_{\rm int} = -U\sum_{\vc{i}\alpha} \hat{c}^\dagger_{\vc{i}\alpha\uparrow}\hat{c}_{\vc{i}\alpha\uparrow}\hat{c}^\dagger_{\vc{i}\alpha\downarrow}\hat{c}_{\vc{i}\alpha\downarrow}$ ($U>0$). Due to translational symmetry ($K^\sigma_{\vc{i}\alpha,\vc{j}\beta} = K^\sigma_{\alpha,\beta}(\vc{i}-\vc{j})$), the single-particle band dispersions $\varepsilon_{n\vc{k}\sigma}$ and periodic Bloch functions $\ket{u_{n\vc{k}\sigma}}$ are obtained from the eigenvalue equation $\widetilde{K}^\sigma(\vc{k})\ket{u_{n\vc{k}\sigma}} = \varepsilon_{n\vc{k}\sigma}\ket{u_{n\vc{k}\sigma}}$, with $\widetilde{K}^\sigma(\vc{k})$ the Fourier transform of the hopping matrix $K^\sigma$~\cite{peotta2015}. We also assume time-reversal symmetry $K^\uparrow_{\vc{i}\alpha,\vc{j}\beta} = (K^{\downarrow}_{\vc{i}\alpha,\vc{j}\beta})^*$ and define $u_{n\vc{k}}(\alpha) \equiv u_{n\vc{k}\uparrow}(\alpha) = u_{n,-\vc{k},\downarrow}^*(\alpha)$, $\varepsilon_{n\vc{k}}\equiv \varepsilon_{n\vc{k}\uparrow} = \varepsilon_{n,-\vc{k},\downarrow}$.  In the following we always use Greek letters $\alpha$, $\beta$ to label the sublattices, while Latin letters $m$, $n$ are band labels. 

From the solution of the two-body problem in a flat band (labeled by $\bar{n}$), it has been shown that the effective mass of the bound state is given approximately by~\cite{torma2018}
\begin{equation}
	\label{eq:eff_mass_flat}
	\begin{split}
	\left[\frac{1}{m_{\rm eff}}\right]_{ij}  &\approx \frac{U\Omega_c}{\hbar^2}\int \frac{d^d\vc{k}}{(2\pi)^d}\,\sum_{\alpha = 1}^{N_{\rm orb}} \partial_{i}P_{\alpha\alpha}(\vc{k})\partial_jP_{\alpha\alpha}(\vc{k}) \\
	&\approx \frac{U\Omega_c}{N_{\rm orb}\hbar^2} \int \frac{d^d\vc{k}}{(2\pi)^d}\,g_{jl}(\vc{k})\,.
	\end{split}
\end{equation}
Here $\Omega_{\rm c}$ is the unit cell volume, $\partial_i \equiv \partial_{k_i}$ and $P_{\alpha\beta}(\vc{k}) = u_{\bar{n}\vc{k}}(\alpha)u^*_{\bar{n}\vc{k}}(\beta)$ are the matrix elements of the band projector $P(\vc{k}) = \ket{u_{\bar{n}\vc{k}}}\bra{u_{\bar{n}\vc{k}}}$ of the flat band, an $N_{\rm orb}\times N_{\rm orb}$ matrix. The quantity $g_{jl}(\vc{k})$ is the quantum metric (see Box 1). Eq.~\eqref{eq:eff_mass_flat} teaches us that the effective mass of a two-body state in a flat band depends  both on the interaction strength and on the quantum metric, a single-particle property.

This result shows that Cooper pairs can have a finite effective mass and thus support transport if the flat band quantum metric is nonzero. This is in sharp contrast with Eq.~\eqref{eq:Ds_BCS_single_band}, which gives zero superfluid weight for a partially filled single (or non-hybridizing) flat band, showing that Eq.~\eqref{eq:Ds_BCS_single_band}
	neglects additional contributions that come into play in the case of multiband/multiorbital lattices ($N_{\rm orb} > 1$). These were identified only recently by applying standard BCS theory to multiband lattice models~\cite{peotta2015,julku2016,liang2017} and are the subject of the next Section.
	
	\subsection{Conventional and geometric contributions of the superfluid weight}
	
	The superfluid weight is a static transport coefficient defined as~\cite{Scalapino1993}
	\begin{equation}
		\label{eq:superfluid_weight}
		D_{{\rm s},jl}  = -\lim_{\vc{q}_\perp \to 0} \chi_{jl}(q_{\parallel}=0,\vc{q}_{\perp},\omega = 0)\,,
	\end{equation}
	where $\chi_{jl}(\vc{q},\omega)$ is a response function, that of the current induced in the system to linear order in the vector potential $\vc{A}$.  The vector potential  enters in the single-particle Hamiltonian $\hat{H}_0$ by the usual Peierls substitution~\cite{Scalapino1993}.
	In the above equation, the wavevector $\vc{q} = q_\parallel \hat{l} + \vc{q}_\perp$ is decomposed into the collinear ($q_{\parallel}$) and perpendicular ($\vc{q}_\perp$) components with respect to the $l = x,y,z$ axis (note that $l$ is the second index appearing in $D_{{\rm s},jl}$ and $\chi_{jl}$). 
	
	To evaluate Eq.~\eqref{eq:superfluid_weight} it is necessary to resort to approximations. The typical BCS approximation is to replace the Hubbard interaction term by~\cite{Schrieffer1964,peotta2015}
	\begin{equation}
		\hat{H}_{\rm int} \approx \hat{H}^{\rm (m.f.)}_{\rm int} = \sum_{\vc{i}\alpha}\left(\Delta_\alpha\hat{c}_{\vc{i}\alpha\uparrow}^\dagger \hat{c}_{\vc{i}\alpha\downarrow}^\dagger + \text{H.c.}\right)\,.  
	\end{equation}
	The pairing field $\Delta_\alpha$ is  independent of the unit cell index $\vc{i}$ since translational symmetry is assumed, and is calculated self-consistently as $\Delta_\alpha = -U\langle \hat{c}_{\vc{i}\alpha\downarrow} \hat{c}_{\vc{i}\alpha\uparrow}  \rangle$, where the expectation value is taken with respect to the mean-field Hamiltonian $\hat{H}_{\rm m.f.} = \hat{H}_0 + \hat{H}_{\rm int}^{\rm (m.f.)}$.
	
	For illustration purposes we consider the case in which time-reversal symmetry is present and the pairing field is real for small vector potentials and orbital-independent $\Delta_\alpha = \Delta$ (see Ref.~\onlinecite{Tovmasyan2016} for a justification, and Ref.~\onlinecite{Huhtinen2021two} for the general case).
	Under these assumptions the superfluid weight can be written as the sum of two terms $D_{s} = D_{\rm conv} + D_{\rm geom}$~\cite{peotta2015,liang2017}. The first term reads
	\begin{gather}
		\label{eq:Ds_conv}
		\begin{split}
			D_{{\rm conv},jl} = \frac{e^2}{\hbar^2}  &\int \frac{d^d\vc{k}}{(2\pi)^d}\,\sum_n\bigg[-\frac{\beta}{2\cosh^2(\beta E_{n\vc{k}}/2)} \\ 
			&+\frac{\tanh(\beta E_{n\vc{k}}/2)}{E_{n\vc{k}}}\bigg]\frac{\Delta^2}{E_{n\vc{k}}^2}\partial_j\varepsilon_{n\vc{k}} \partial_{l}\varepsilon_{n\vc{k}}\,,
		\end{split}
	\end{gather}
	and is called the conventional contribution to the superfluid weight since it is the sum over all bands of the result for the superfluid weight of a single-band model, Eq.~\eqref{eq:Ds_BCS_single_band} (now at finite temperature $T$, $\beta^{-1} = k_{\rm B}T$, $k_{\rm B}$ the Boltzmann constant). In Eq.~\eqref{eq:Ds_conv}, $E_{n\vc{k}} = \sqrt{(\varepsilon_{n\vc{k}}-\mu)^2+\Delta^2}$ are the quasiparticle excitation energies of $\hat{H}_{\rm m.f.}$.
	
	The second term is called the geometric contribution to the superfluid weight
	\begin{gather}
		\label{eq:Ds_geom}
		\begin{split}
			D_{{\rm geom},jl} &= \frac{e^2\Delta^2}{\hbar^2}  \int \frac{d^d\vc{k}}{(2\pi)^d}\,\sum_{n\neq m} \bigg[\frac{\tanh(\beta E_{n\vc{k}}/2)}{E_{n\vc{k}}}\\ &-\frac{\tanh(\beta E_{m\vc{k}}/2)}{E_{m\vc{k}}}\bigg]\frac{(\varepsilon_{n\vc{k}}-\varepsilon_{m\vc{k}})^2}{E_{m\vc{k}}^2-E_{n\vc{k}}^2}\\ &\times
			\big(\dbk{\partial_j u_{n\vc{k}}}{u_{m\vc{k}}} \dbk{u_{m\vc{k}}}{\partial_l u_{n\vc{k}}}+(j\leftrightarrow l)\big)\,.
		\end{split}
	\end{gather}
	This contribution is a qualitatively new feature of multiband/multiorbital lattice models and is associated to off-diagonal matrix elements of the current operator, which are proportional to
	\begin{equation}
		\label{eq:dk_K}
		\begin{split}
			\bra{u_{m\vc{k}}}\bm{\nabla}_{\vc{k}}\widetilde{K}^\uparrow(\vc{k})&\ket{u_{n\vc{k}}} = \bm{\nabla}_{\vc{k}}\varepsilon_{n\vc{k}}\delta_{n,m} \\ &+  (\varepsilon_{n\vc{k}}-\varepsilon_{m\vc{k}})\langle u_{m\vc{k}} | \bm{\nabla}_{\vc{k}}u_{n\vc{k}} \rangle\,. 
		\end{split}
	\end{equation}
	Indeed, the diagonal term $\propto \bm{\nabla}_{\vc{k}}\varepsilon_{n\vc{k}}$ leads to the conventional contribution Eq.~\eqref{eq:Ds_conv} and the off-diagonal term $\propto \langle u_{m\vc{k}} | \bm{\nabla}_{\vc{k}}u_{n\vc{k}} \rangle$ to the geometric one Eq.~\eqref{eq:Ds_geom}. Further details are given in Refs.~\cite{liang2017,peotta2015,Huhtinen2021two} and the review Ref.~\cite{Rossi2021}.
	
	It is instructive to investigate the two contributions Eqs.~\eqref{eq:Ds_conv}-\eqref{eq:Ds_geom} in the isolated band limit, when the chemical potential is such that a single band, denoted by $\bar{n}$, is partially filled and other bands are separated from it  by a large band gap $E_{\rm gap}$
	\begin{equation}
		\label{eq:isol_band_lim}
		|\varepsilon_{m\vc{k}}-\varepsilon_{\bar{n}\vc{k}}| \gtrsim E_{\rm gap} \gg \Delta \quad \text{for} \quad m\neq \bar{n}\,.
	\end{equation}
	The conventional contribution in this case is just Eq.~\eqref{eq:Ds_BCS_single_band} with $\varepsilon(\vc{k})$ replaced by $\varepsilon_{\bar{n}\vc{k}}$. On the other hand, care is required  for the computation of the geometric contribution. One might expect that the geometric contribution should be vanishing in this limit since the terms of the response function that are off-diagonal in band space are suppressed by the factor $E_{m\vc{k}}-E_{\bar{n}\vc{k}} \gtrsim E_{\rm gap}$, with $m\neq \bar{n}$, at the denominator. This would be incorrect because the off-diagonal terms of the current operator are proportional to the band gap, as seen from Eqs.~\eqref{eq:dk_K}-\eqref{eq:isol_band_lim}. A careful analysis~\cite{liang2017} shows that the geometric contribution is nonzero even in the isolated band limit and, somewhat counter-intuitively, can be expressed purely in terms of quantities relative to the partially filled band, in particular its quantum metric. This is most clearly illustrated in the case of an isolated flat band, where one obtains ($\nu$ is the band filling)~\cite{peotta2015}
	\begin{equation}
		\label{eq:Ds_flat}
		D_{{\rm s},jl} =D_{{\rm geom},jl} = \frac{4e^2\Delta\sqrt{\nu(1-\nu)}}{\hbar^2}\int \frac{d^d\vc{k}}{(2\pi)^d}g_{jl}(\vc{k})\,,
	\end{equation}
	which relates the superfluid  weight to the quantum metric  $g_{jl}(\vc{k})$ integrated over the BZ. In the general case, when $\Delta$ is not necessarily real for all vector potentials, $g_{jl}$ is replaced by the minimal quantum metric~\cite{Huhtinen2021two}. It is clear by comparing Eq.~\eqref{eq:Ds_flat} with Eq.~\eqref{eq:eff_mass_flat} that, with some minor approximations, the nonzero superfluid weight in the flat band limit can be entirely explained in terms of the two-body bound state effective mass, Section~\ref{section:pairmass}~\cite{torma2018,Iskin2021twobody}. The geometric contribution scales with the superconducting energy gap $\Delta \propto U$, while the conventional one scales roughly with the Fermi energy $E_{\rm F}$, which is of the order of the bandwidth of the partially filled band. In usual, non-flat band, superconductors the gap $\Delta$ is generally a much smaller energy scale than the Fermi energy and the geometric contribution is rather small.
	
	Only very recently physical systems in which the electronic bandwidth is comparable to the superconducting gap have been realized in the laboratory, in particular magic-angle TBG \cite{2021YazdaniGap}. The essential role of the off-diagonal matrix elements of the current operator  was pointed out for the first time in  Ref.~\onlinecite{Moon1995} in the case of the exciton superfluid phase occuring in quantum Hall bilayers, which can be considered the first example of a flat band superfluid observed experimentally. Moreover, Kopnin~showed theoretically that the superfluid weight is nonzero in the case of the flat band of surface states present in rhombohedral graphene~\cite{Kopnin2011a}. 
	However, these have apparently been regarded as system-specific findings until general results connecting superfluidity with quantum geometry, such as Eqs.~\eqref{eq:Ds_geom} and~\eqref{eq:Ds_flat}, were obtained~\cite{peotta2015,liang2017}.

	\subsection{Flat band superconductivity and Wannier function overlap} \label{Wfunctionoverlap}
	
	A natural question raised by Eq.~\eqref{eq:Ds_flat} is what property of the flat band wave-functions is captured by the quantum metric and why it affects transport and superfluidity. The physical interpretation of the quantum metric can be most easily provided by Wannier functions,
	that is, Fourier transforms of the Bloch states
	\begin{equation}
	w_n(\vc{i},\alpha) =\frac{\Omega_{\rm c}}{(2\pi)^2}\int d^d\vc{k}\,e^{i\vc{k}\cdot\vc{r}_{\vc{i}\alpha}}u_{n\vc{k}}(\alpha). 
	\label{eq:Wannierdefinition}
	\end{equation}
	A common prescription to choose a gauge for the Wannier functions is the one based on the Marzari-Vanderbilt localization functional whose minimization leads to the maximally localized Wannier functions~\cite{Marzari2012}. The gauge-invariant part of the localization functional is the trace of the quantum metric integrated over the BZ, which sets a lower bound on how much the Wannier functions can be localized. It was shown in Ref.~\onlinecite{Tovmasyan2016} that a flat band superconductor can be described by an effective spin Hamiltonian whose exchange couplings are controlled by the overlap of the Wannier functions, which in turn is related to the Marzari-Vanderbilt functional and thus to the quantum metric. This provides a solid basis for the interpretation of the quantum metric as an invariant measure of the overlap/spread of the flat band wave functions.
	
	Topological invariants that describe the electronic band structure provide an obstruction to the full localization of Wannier functions. An important topological invariant is the Chern number $\mathcal{C}$~\cite{Chiu2016}, see Box 1. From the positive semi-definiteness of the QGT one obtains the inequality
	\begin{equation}
		\label{eq:ineq}
		\mathrm{det}\,\mathcal{M}^{\rm R} \geq \mathcal{C}^2\,,\,\,\text{with}\,\,\mathcal{M}^{\rm R}_{ij} =\frac{1}{2\pi}\int d^2\vc{k}\,g_{ij}(\vc{k})\,. 
	\end{equation}
	A nonzero Chern number hence also bounds the superfluid weight, since the matrix $\mathcal{M}^{\rm R}$ is precisely the integrated quantum metric that enters in Eq.~\eqref{eq:Ds_flat}. This is explained by the fact that if the Chern number is nonzero one cannot find exponentially localized Wannier functions~\cite{Brouder2007,Panati2007}; at best the Wannier states are algebraically decaying with a known exponent $1/r^2$~\cite{Monaco2018}. Such wave functions are necessarily overlapping and thus are expected to support a robust  superfluid state in the isolated flat band limit. The Marzari-Vanderbilt localization functional and the quantum metric appear also in an upper bound for the superfluid weight, derived without the use of a mean-field approximation~\cite{Verma2021}.
	
	Inequalities similar to Eq.~\eqref{eq:ineq} have been discovered in the case of two more topological invariants: the winding number in $d=1$~\cite{Tovmasyan2016}, and the Euler class in $d = 2$~\cite{xie2020} which is relevant for magic angle TBG and will be discussed in detail in Sections~\ref{sec:fragile_topology} and~\ref{sec:lower_bounds_TBG}. One expects such bound identities to exist for every topologically nontrivial band. Topology is a sufficient but not necessary condition for non-zero superfluid weight~\cite{julku2016}, because the quantum metric can be non-zero for zero Chern number, even zero Berry curvature bands -- a bound that describes also such bands was recently defined using obstructed Wannier centers~\cite{Herzog-Arbeitman2021}.

	\subsection{Berezinskii-Kosterlitz-Thouless temperature of superconductivity in a flat band}
	\label{sec:BKT_flat _band}
	
	The superfluid weight controls the susceptibility of the order parameter phase to thermal fluctuations, as seen from Eq.~\eqref{eq:free_energy_Ds}, and thus it is expected to affect the critical temperature of superconductivity. However, this effect is not captured by BCS theory in its simplest form since 
	the order parameter $\Delta(\mathbf{r})$ is a non-fluctuating quantity at the mean-field level. In $d=2$ the relation between critical temperature and superfluid weight can be made precise thanks to the universal relation $T_{\rm BKT} = \frac{\pi \hbar^2}{8e^2}D_{\rm s}(T_{\rm BKT})$~\cite{Nelson1977}, where $T_{\rm BKT}$ is the Berezinskii–Kosterlitz–Thouless (BKT) temperature, that is the critical temperature  of a two-dimensional superfluid, and $D_{\rm s}(T_{\rm BKT})$ is the superfluid weight slightly below the same temperature.
	For an isolated flat band one obtains that $T_{\rm BKT}\sim U$~\cite{peotta2015,liang2017}, that is, the same linear dependence on $U$ as for the BCS critical temperature in a flat band. The BKT temperature saturates for an interaction strength of the order of the band gap $U\sim E_{\rm gap}$ and then decreases as $U^{-1}$ in the strong-coupling limit. This behavior can be understood with a strong-coupling expansion in which the Cooper pairs are tightly bound and move in the lattice with an effective hopping $J \sim t^2/U$, with $t$ the scale of the single particle hopping~\cite{liang2017}. The linear dependence of the critical temperature and of the superfluid weight with the interaction strength for $U\lesssim E_{\rm gap}$ is a robust result that has been verified using advanced many-body methods such as dynamical mean-field theory~\cite{julku2016,liang2017}, quantum Monte Carlo~\cite{Hofmann2020,Peri2021}, and density matrix renormalization group~\cite{Tovmasyan2018,Mondaini2018,Chan2021}. The dependence can be almost linear also in the presence of a band touching between the flat band and other bands~\cite{julku2016}.

\section{Superfluidity and topology in twisted bilayer graphene (TBG)}

Remarkable interacting phases have been observed in TBG when the two layers are twisted with respect to each other by a "magic" angle $\theta\approx 1.1^\circ$, see Fig.~\ref{MagneticSGandTQC1}a. These phases include correlated (sometimes Chern) insulators at integer filling $\nu$ of a set of 8 (2 per valley per spin quantum numbers) flat bands around charge neutrality, with superconductors for fillings in between \cite{bistritzer_moire_2011,cao_correlated_2018,cao_unconventional_2018, lu2019superconductors, yankowitz2019tuning, saito_independent_2020, stepanov_interplay_2020, liu2020tuning, serlin_QAH_2019,  jiang_charge_2019,xie2019spectroscopic, choi_imaging_2019,  nuckolls_chern_2020, choi2020tracing,das2020symmetry, wu_chern_2020, lu2020fingerprints,   burg2020evidence, zou2018, fu2018magicangle, liu2019pseudo, kang_symmetry_2018, song_all_2019,po_faithful_2019,ahn_failure_2019,Slager2019WL, lian2020, hejazi_landau_2019, kang_strong_2019,bultinck_ground_2020,po_origin_2018,xie2020,julku2020, hu2019, kang_nonabelian_2020, soejima2020efficient, pixley2019, xie_HF_2020,liu2020theories, cea_band_2020,daliao2020correlation,abouelkomsan2020,repellin_FCI_2020, vafek2020hidden, fernandes_nematic_2020,  Wilson2020TBG,wang2020chiral, ourpaper2,ourpaper5,Codecido2019}, see Fig.~\ref{MagneticSGandTQC1}b.  Scanning tunneling microscopy (STM) experiments reveal a Coulomb repulsion strength ($\sim 25$meV) \cite{xie2019spectroscopic} larger than the electron bandwidths. The strongly correlated insulators  \cite{xie2019spectroscopic} at integer filling $\nu \in [-4,4]$  measured from charge neutrality acquire Chern numbers $\pm(4-|\nu|)$ \cite{nuckolls_chern_2020,choi2020tracing,serlin_QAH_2019,das2020symmetry, wu_chern_2020}.  Theoretically, the interacting TBG~\cite{roy:2018,kang_strong_2019,bultinck_ground_2020, po_origin_2018,xie2020,julku2020, hu2019, kang_nonabelian_2020, soejima2020efficient, pixley2019,  xie_HF_2020,liu2020theories, cea_band_2020,daliao2020correlation,abouelkomsan2020,repellin_FCI_2020, vafek2020hidden, fernandes_nematic_2020,wang2021exact}, is governed by \cite{kang_strong_2019}  an approximate U(4) symmetry (later extended to U(4) $\times$ U(4)~\cite{kang_symmetry_2018,vafek2020hidden}). The correlated insulator phases can be understood as ferromagnets of this large symmetry group~\cite{kang_strong_2019, ourpaper4,bultinck_ground_2020, vafek2020hidden}.

\subsection{TBG superconductivity}

Moir\'e materials host superconductivity \cite{ cao_unconventional_2018, yankowitz2019tuning,stepanov_interplay_2020,Zhang2021WangGroup,chen_signatures_2019}  even at the -- by far -- lowest densities ever: $10^{10} - 10^{12} cm^{-2}$ 
\cite{chen_signatures_2019,Zhang2021WangGroup}, which for the sample sizes used means 1000-10000 electrons. What causes $\approx 1-5\mathrm{K}$ superconductivity in this most dilute of all known systems? Experimental evidence links superconductivity in TBG  to the presence of flat bands at the magic angle. Away from these angles, superconductivity disappears.  Regardless of  the mechanism of TBG superconductivity, since the TBG flat bands exhibit zero total Chern number, is there a bound on their superfluid weight? What influence could the quantum geometry of the TBG bands have on the superfluid weight of TBG? 

\begin{figure*}
\centering
\includegraphics[height=4.5in, width=6.7in]{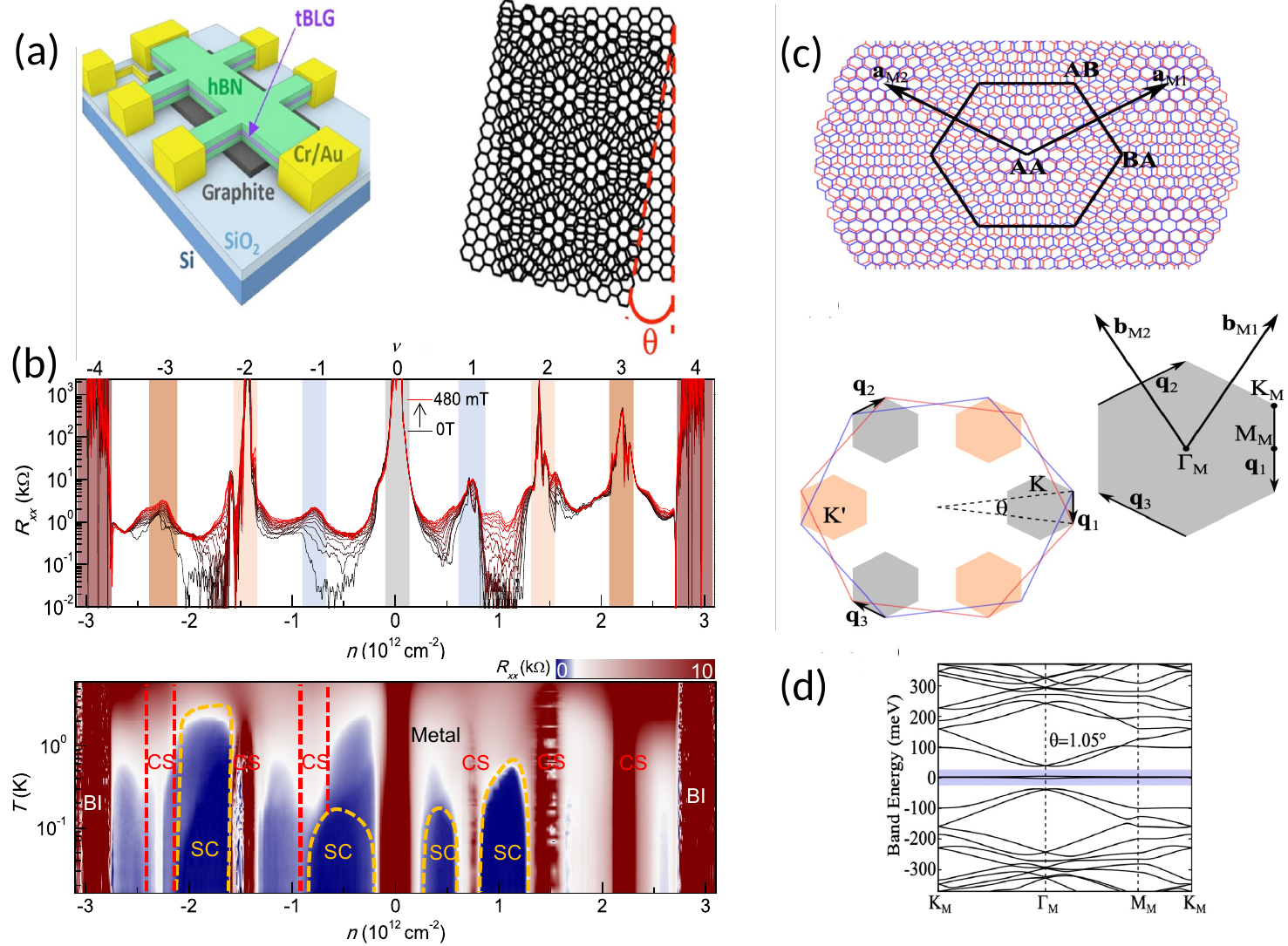}
\caption{\textbf{TBG lattice, experimental phase diagram, and band structures.} 
\textbf{(a)} Sample for encapsulated \MR~TBG (tBGL in the picture) lattice for a rotation angle $\theta$. \textbf{(b)} Experimental observation of correlated insulators (correlated states, CS) -- the red regions of the plots where the resistance ($R_{xx}$) peaks -- at integer fillings $\nu$ of the two flat bands shaded in (d), and superconducting (SC) phases (low resistance regions in blue) in between the insulating phases, and band insulators (BI). Here $T$ is the temperature and $n$ is the density. Obtained with permission from~\cite{lu2019superconductors}. \textbf{(c)} Top: The \MR~unit cell, where the blue sheet and the red sheet represent the top and bottom layers, respectively. In the AA, AB, BA regions, the A sublattice of the top layer are located above the A sublattice, the B sublattice, and the hexagon center of the bottom layer, respectively. Bottom: the \MR~BZ. Left: The grey and yellow hexagons represent the \MR~BZ for the graphene valleys $K$ and $K'$, respectively. Right: The reciprocal lattices and the high symmetry momenta of the \MR~BZ in graphene valley $K$. \textbf{(d)} The band structure of the magic-angle $\theta=1.05^\circ$ TBG with its two flat bands (for one valley) using the BM model. Panel (a) adapted from Ref.~\cite{Codecido2019}, under a Creative Commons licence CC BY-NC 4.0. Panel (b) adapted from Ref.~\cite{lu2019superconductors}, Springer Nature Ltd. Panels (c) and (d) adapted with permission from Ref.~\cite{ourpaper2}, APS.
}\label{MagneticSGandTQC1}
\end{figure*}

The moir\'e unit cell contains tens of thousands of atomic sites for small angles; such a large number of bands is unpractical to handle in most calculations. Fortunately, many approximate approaches exist. At low energies, the single particle TBG flat bands are well described by the Bistritzer-MacDonald (BM) model~\cite{bistritzer_moire_2011}. This model describes the coupling of the band structures of the top and bottom layers, which are rotated with respect to each other by $\pm \theta/2$ around the $z$-axis. Each band structure separately consists of two Dirac points at the time-reversal ($\TRS$) partner momenta $K$ and $K'$ of the single layer graphene BZ (for illustrations of the BZs see Fig.~\ref{MagneticSGandTQC1}c). When $\theta$ is small such that the interlayer coupling is smooth in real space (with a length scale of variation much larger than the atom distances), the graphene valley ($K$ and $K'$) is a good quantum number of low energy states of TBG \cite{bistritzer_moire_2011} and due to its conservation the Dirac states around $K$ ($K'$) in the top layer only couple to the states around $K$ ($K'$) in the bottom layer. The Dirac Hamiltonian around $K$ in the top layer is $-i v_F\pt_x (\cos\frac{\theta}2\sigma_x-\sin\frac{\theta}2\sigma_y) -i v_F \pt_y (\cos\frac{\theta}2\sigma_y+\sin\frac{\theta}2\sigma_x) \approx -iv_F \pt_\rr\cdot\bsigma + i\frac{\theta}2 v_F \pt_\rr\times \bsigma$, where $v_F$ is the Fermi-velocity of single-layer graphene and $\bsigma=(\sigma_x,\sigma_y)$ are Pauli matrices representing the A/B sublattices of graphene.
For the bottom layer we replace $\theta \rightarrow -\theta$. The interlayer coupling is encoded in a position-dependent matrix $T(\rr)$, such that the Hamiltonian of TBG, at valley $K$, can be written as 
\begin{equation}
H(\rr) = 
-iv_F \tau_0 \pt_\rr\cdot\bsigma - \frac{\theta}2 \tau_z \pt_\rr\times \bsigma +
\begin{pmatrix}
 0  &  T(\rr)  \\
 T^\dagger(\rr) &  0 
\end{pmatrix} . \label{eq:Ham}
\end{equation}Here $\tau_0$ and $\tau_z$ are the identity and the third Pauli matrix for the layer degree of freedom, respectively.
 Ref.~\cite{bistritzer_moire_2011} derived  $T(\rr)$ for  small $\theta$ $\sim 1^\circ$:
\begin{equation}
    T(\rr) = \sum_{i=1}^3 e^{-i\qq_i \cdot \rr} T_i, \label{eq:Tr}
\end{equation}
where $\qq_i$'s are $\qq_1=k_D(0,-1)$, $\qq_2=k_D(\frac{\sqrt3}2,\frac12)$, $\qq_3=k_D(-\frac{\sqrt3}2,\frac12)$, with $k_D=2|K|\sin\frac{\theta}2$ being the distance between $K$ momenta in the two layers, and $v_F=5.944{\rm eV \cdot \mathring{ A} }$, $|K|=1.703\mathring{\rm A}^{-1}$. The $T_i$'s are 
\begin{equation} \label{eq:Ti-def}
\begin{aligned}
    T_i =& w_0 \sigma_0 + w_1\Big[\sigma_x\cos\frac{2\pi(i-1)}3+ \sigma_y\sin\frac{2\pi(i-1)}3\Big],
\end{aligned}
\end{equation}
where the AA and AB hopping are  $w_1=110{\rm meV}$, $w_0=0.7w_1$, respectively. The translation symmetry of the \MR~potential Eq.~(\ref{eq:Tr}) under the \MR~unit cell translations $\aa_{M1} = \frac{2\pi}{k_D} (\frac1{\sqrt3}, \frac13)$,
$\aa_{M2} = \frac{2\pi}{k_D} (-\frac1{\sqrt3}, \frac13)$ is manifest in real space, see Fig.~\ref{MagneticSGandTQC1}c.
At $\theta \approx 1.1^\circ$, two very flat bands can be seen in  Fig.~\ref{MagneticSGandTQC1}d (marked with blue); these are conventionally referred to as the "active bands" of TBG.

\subsection{The BKT temperature and geometric contribution in TBG} \label{BKTTemp}

Being near the BKT transition is manifested in electron transport as $V \propto I^3$ dependence between the longitudinal voltage $V$ and the current $I$. Using this, $T_{\rm BKT}$ was estimated in Ref.~\cite{cao_unconventional_2018} to be about 1 K, however, the fit to $I^3$ was partial. In a more recent work on magic-angle twisted \textit{trilayer} graphene, a clear $I^3$ behaviour was observed, giving $T_{\rm BKT} \simeq 2.1 $K~\cite{Park2021}.  

\begin{figure}[h]
\includegraphics[width=1.0\columnwidth]{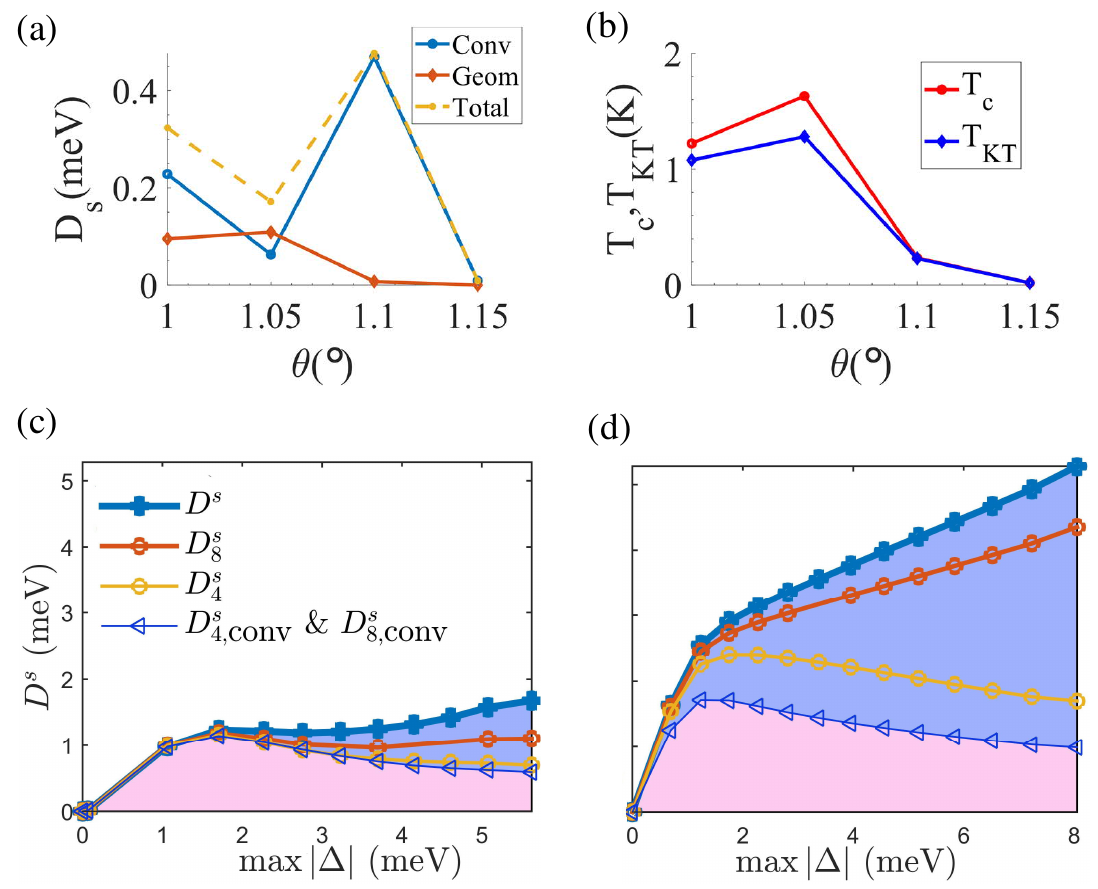}
\caption{\textbf{Geometric contribution in TBG superfluid weight.} Superfluid weight $D_s$ \textbf{(a)} and the critical temperature \textbf{(b)} calculated in~\cite{hu2019}, as function of the twist angle $\theta$. The geometric contribution of $D_s$ is larger than the conventional one at the magic angle 1.05$^\circ$ while away from it the conventional one dominates; the critical temperature changes from BKT type ($T_{\rm BKT}$) to the conventional BCS one ($T_{\rm c}$) when moving away from the magic angle. \textbf{(c-d)} Superfluid weight as a function of the order parameter (as it is spatially varying, the maximum $|\Delta|$ is used), from~\cite{julku2020}. The order parameter is varied by changing the coupling constant of the interaction term, which is a nearest neighbor (RVB) interaction in \textbf{(c)} and a local interaction in \textbf{(d)}. Here $D_s^{4}$ and $D_s^{8}$ are the superfluid weights calculated including only the four nearly flat bands, or the eight lowest (four flat, four dispersive) bands, respectively, and $D_s$ is the converged result including 40 bands of the 676 of the RM model used in~\cite{julku2020}. For interactions large compared to the bandwidth, the flat band limit is reached and the geometric contribution (blue area) becomes larger than the conventional one (pink). The lines marked with different symbols show that the conventional contribution is well captured by including only the four lowest bands in the calculation, while more than eight are needed to obtain the correct result in the case of the geometric contribution which arises from interband processes. Figures obtained from~\cite{hu2019} and~\cite{julku2020} with permission, APS.}
\label{fig:BKTtemperature}
\end{figure}

Theoretically, the BKT temperature of TBG superconductors was estimated for the first time in Refs.~\cite{hu2019,julku2020,xie2020}; for a brief introduction see~\cite{classen2020}. Ref.~\cite{hu2019} utilized the low-energy BM model described above and considered local $s$-wave pairing leading to a local mean-field order parameter $\Delta$. Ref.~\cite{julku2020} presents a more microscopic approach where the full tight-binding TBG Hamiltonian is reduced to a model that is manageable but still has a large unit cell of a few hundreds of atoms, by using the so-called renormalization moir\'e (RM) method~\cite{su:2018,lopesdossantos:2007}. A low-energy continuum calculation was presented as a comparison to the RM calculations. Both local $s$-wave and nearest-neighbour resonance valence bond (RVB) pairing were considered, in the latter the order parameter $\Delta$ corresponds to pairing of electrons in neighbouring sites.   
Both works~\cite{hu2019} and~\cite{julku2020} found that $T_{\rm BKT}$ at the magic angle is significantly influenced by the geometric contribution of superconductivity Eq.~\eqref{eq:Ds_geom}. In Ref.~\cite{hu2019} it was shown that the geometric contribution to the superfluid weight is about twice as large than the conventional one at the magic angle, while slightly away from it the conventional one starts to dominate, see Fig.~\ref{fig:BKTtemperature}(a-b). As the pairing mechanism and the interaction strength in TBG are not precisely known at the moment, no accurate predictions of $T_{\rm BKT}$ could be made in Refs.~\cite{hu2019,julku2020,xie2020}, however, the estimates gave temperatures in the few Kelvin range, consistent with the experiments. Assuming $\Delta(T) \approx 2k_BT_c^*(1-T/T_c^*)^{1/2}$ \cite{Tinkham2004}, the ratio between the Cooper pair formation (BCS) critical temperature $T_c^*$ and $T_{\rm BKT}$ was found for the filling ratio $\nu=1/4$ (2 electrons per moir\'e unit cell) to be $T_{\rm BKT}/T_c^*=0.35$~\cite{xie2020}. For the geometric contribution of the TBG superfluid weight $D_s$ at zero temperature, using the experimental $T_{\rm BKT} = 1.5 \rm\,K$ \cite{cao_unconventional_2018} and an order parameter $\Delta = 2k_B T_c^* \approx 0.74 \rm\,meV$, the value $[D_s]_{\rm geom} \approx \frac{4e^2\Delta}{\pi \hbar^2}\sqrt{\nu(1-\nu)} \approx 1.5 \times 10^8 \rm \, H^{-1}$ was obtained~\cite{xie2020,julku2020,hu2019}. This is \textit{one} order of magnitude smaller than the superfluid weight in BSCCO and MoGe,  
but for a TBG density \textit{two} orders of magnitude smaller ($n\approx 10^{12}\,\rm cm^{-2}$).

TBG superconductivity is often described by models featuring only the four nearly flat bands, as it is assumed that the low-energy physics dominates. Importantly, however, the work~\cite{julku2020} suggests that neglecting higher bands may not be adequate when calculating the BKT temperature;
the underlying reason is that the geometric contribution of the superfluid weight is due to interband processes which do \textit{not} scale as one over the band gap as one might naively expect for processes involving higher bands (see Eqs.~\eqref{eq:Ds_geom}, \eqref{eq:dk_K}, and Ref.~\cite{liang2017}). Fig.~\ref{fig:BKTtemperature}(c-d) shows how the TBG superfluid weight divides into the conventional and geometric contributions, the latter becoming dominant when interactions exceed the bandwidth, that is, the flat band regime is approached. Fig.~\ref{fig:BKTtemperature}(c-d) makes clear that considerably more than four bands need to be included to get quantitatively or even qualitatively correct results. In choosing effective low-energy models with only a few bands, one should thus pay attention to the quantum geometric properties of its Bloch functions, in addition to the dispersion and symmetries. One more finding in~\cite{julku2020} was that, in case of the RVB pairing, the superfluid is nematic and the corresponding superfluid weight anisotropic – thus measurements of the superfluid weight could distinguish between local or non-local pairing mechanism through (an)isotropy. 

The linear dependence of the BKT temperature on interaction in case of TBG was confirmed by Monte Carlo calculations~\cite{Peri2021}. The same was found to hold for the temperatures where the single-particle density of states and the spin susceptibility reach their maxima. These temperatures were higher than the BKT temperature, indicating singlet and gap formation already above the temperature for superconductivity. Such pseudogap formation was also studied in Ref.~\cite{WangLevin2020} for a flat band model different from TBG. It was pointed out that the ratio between the pseudogap temperature and $T_{\rm BKT}$ can tell about the geometric contribution of the preformed pair effective mass.  

Besides \MR{} materials, the geometric contribution to the superfluid weight has been found to be important in another 2D superconductor, monolayer $\mathrm{FeSe}$ grown on a $\mathrm{SrTiO}_3$ (STO) substrate~\cite{Kitamura2021}, which has a critical temperature of $65\mathrm{K}$. $\mathrm{FeSe}$ on STO has a large ratio $T_{\rm c}/T_{\rm F} \sim 0.1$ between critical temperature  and Fermi temperature, even larger than that of TBG~\cite{cao_unconventional_2018} (a property shared with its parent compound: bulk $\mathrm{FeSe}$~\cite{Lee2017}). These findings  validate the general picture that the effect of quantum geometry on superfluidity is crucial in materials close to the flat band limit.

\subsection{Symmetries of TBG, Fragile and stable topology}
\label{sec:fragile_topology}

\begin{figure*}
\centering
\includegraphics[height=4.5in, width=6.7in]{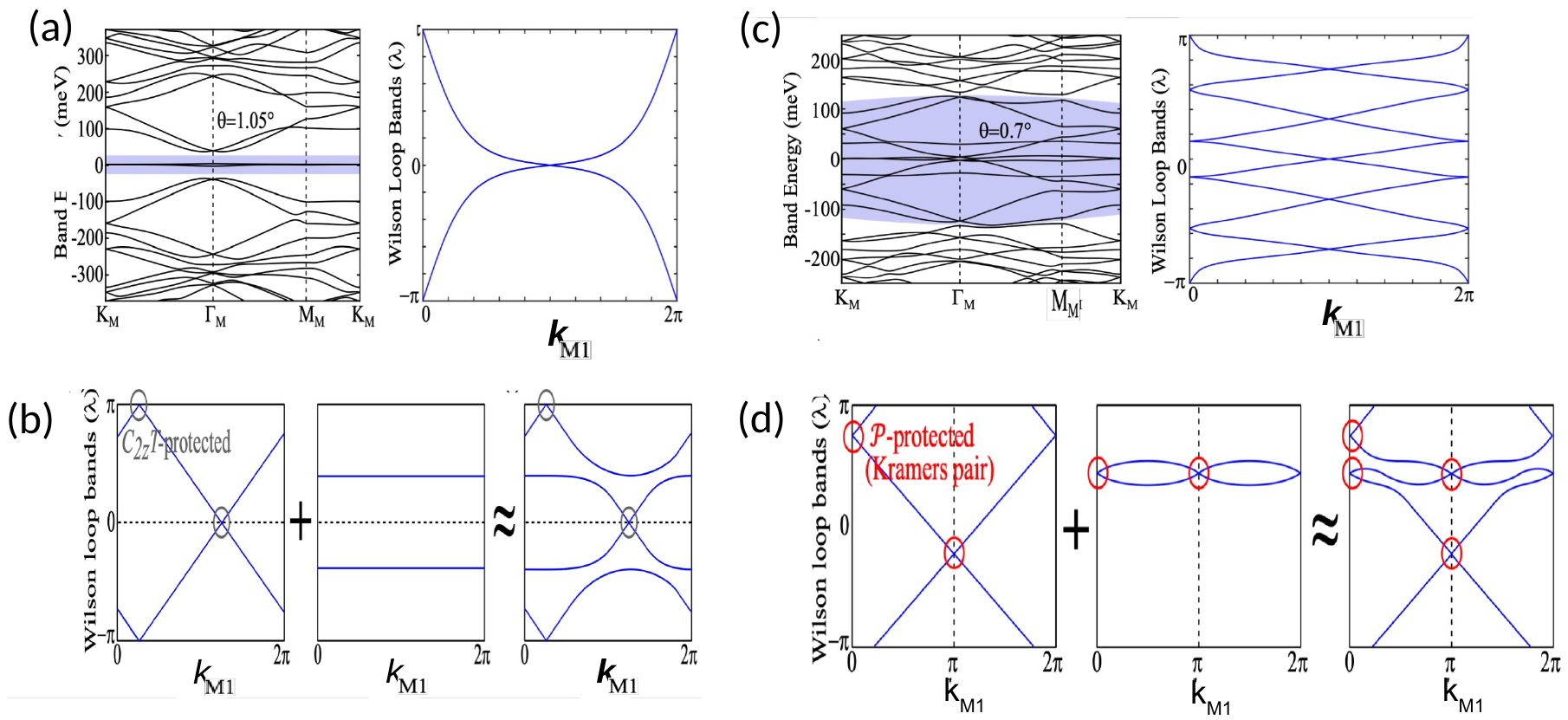}
\caption{\textbf{Wilson loops/Wannier centers, fragile and stable topology of TBG.} 
 \textbf{(a)} Left: The band structure of the magic-angle $\theta=1.05^\circ$ TBG with its two flat bands (using the BM model, single valley). Right: their hybrid Wannier centers -- the Wilson loop bands. The crossings in the Wilson loop bands are protected by $C_{2z}\TRS$ and/or the approximate $\PH$ symmetries. Each Wilson loop operator is integrated along $\bb_{M2}$ (definition in Fig.~\ref{MagneticSGandTQC1}(c)) and the spectrum is plotted along $\bb_{M1}$. \textbf{(b)} The Wilson loop bands with $C_{2z}\TRS$. The crossings at $\lambda=0,\pi$ are protected by $C_{2z}\TRS$. For a two-band system, the topology is nontrivial if the Wilson loop bands -- representing the position of the Wannier center -- wind (left) \emph{across} the unit cell and is trivial otherwise (middle and right). The $C_{2z}\TRS$-protected topology is fragile: Coupling the nontrivial Wilson loop bands (left) to a trivial Wilson loop bands (middle) yields a trivial Wilson loop bands (right), as the bands do not wind over the whole BZ any more. \textbf{(c)} The band structure (left) and Wilson loop bands (right) of the middle ten bands of TBG (shaded) at $\theta=0.7^\circ$. The crossings at the Wilson eigenvalue $\lambda=0,\pi$ in the Wilson loop bands are protected by $C_{2z}\TRS$ and/or by particle-hole $P$; the double degeneracies with $\lambda\neq 0,\pi$ at $k_1=0,\pi$ are protected by the approximate $\PH$ symmetry. These double-degeneracies guarantee winding of the Wilson loops across the BZ for any bands with $4n+2$ Dirac nodes at zero energy. The parameters of the
Hamiltonian used in (a-d) are $v_F=5.944{\rm eV \cdot \mathring{ A} }$, $|K|=1.703\mathring{\rm A}^{-1}$, $w_1=110{\rm meV}$, $w_0=0.7w_1$. \textbf{(d)} Comparison of Wilson loop windings protected by $\PH$ and $C_{2z}\TRS$. The crossings at $k_{M1}=0,\pi$ are Kramers pairs protected by $\PH$. The ${Z}_2$ Euler invariant equals to 1 if the Wilson loop bands form a zigzag connection between $k_{M1}=0$ and $k_{M1}=\pi$ and equals to 0 otherwise. The $\PH$-protected topology is stable against adding trivial bands: Coupling the nontrivial Wilson loop bands (left) to trivial Wilson loop bands (middle) yields nontrivial Wilson loop bands (right). Panels (a)-(d) adapted with permission from Ref.~\cite{ourpaper2, APS.} 
}\label{TBGnewfig}
\end{figure*}

The TBG eigenstates exhibit subtle  topology of either a  "fragile" or stable type depending on the symmetries kept in the model.  The single-graphene-valley Hamiltonian Eq.~\eqref{eq:Ham} has the symmetries of the magnetic space group  $P6'2'2$~\cite{song_all_2019} (\#177.151 in the BNS setting \cite{Bilbao-MSG}): (i) $C_{2z}\TRS=\sigma_x K$, where $K$ is the complex conjugation,
(ii) $C_{3z}=e^{i\frac{2\pi}3\sigma_z}$,
(iii) $C_{2x}=\tau_x\sigma_x$.
 $C_{2z}$ rotation and the time-reversal $\TRS$ separately are not symmetries of Eq.~\eqref{eq:Ham} since they map the graphene valley $K$ into $K'$. A  non-crystalline important symmetry for a small twist angle is  a unitary particle-hole inversion operation $P=i\tau_y$, for which $P H (\rr) P^\dagger = - H(-\rr)$. 
 Crucially, Ref.~\cite{song_all_2019} proved that these symmetries force the single-valley TBG model to be topological at any \emph{small} twist angle and that  
 the  active bands of TBG \emph{cannot} be described by localized symmetric Wannier states. To explain that, remind that the Wannier functions $w_n(\vc{i},\alpha)$, Eq.~\eqref{eq:Wannierdefinition}, are spread in space ($\vc{i}$, $\alpha$ are the unit cell and orbital indices). One can define the \qql center\qqr of the Wannier function as the average location, measured from the origin; scaled by the unit length, this  corresponds to the Berry phase~\cite{Aris2014WL,resta2011}. If one Fourier transforms (Eq.~\eqref{eq:Wannierdefinition}) with $k_y$ only, the Wannier center in the $y$ direction will depend on $k_x$. This motion of the Wannier center as a function of the BZ momentum $k_x \in [0, 2\pi)$ is called the Wilson loop~\cite{Aris2014WL,resta2011} and is shown for TBG in Fig.~\ref{TBGnewfig}a. Importantly, it winds between $\pi$ and $-\pi$. This means that the Wannier functions cannot be simultanously localized in $x$ and $y$: such localization would imply the $y$ Wannier center cannot move across the entire $y$-unit cell as $k_x$ is varied across the BZ.   

The type of topology that the valley-filtered bands respect has been subject to revision. In the initial papers \cite{zou2018, song_all_2019,po_faithful_2019, ahn_band_2018, Slager2020Euler, kang_symmetry_2018} the topology was thought to be of a new, "fragile" type \cite{po_fragile_2018, cano_fragile_2018,Slager2019WL} protected by  $C_{2z}\TRS$ (this type of topology does not require any of the other symmetries of TBG). Fragile topology is an obstruction to constructing symmetric Wannier functions for a given set of bands which  can be \textit{removed} by adding other topologically \textit{trivial}
bands to the Hilbert space \cite{po_fragile_2018, cano_fragile_2018}. 
Computing the Wilson loop of more than the two bands around charge neutrality, Fig.~\ref{TBGnewfig}a, causes crossings at different Wilson eigenvalues (schematically  shown in Fig.~\ref{TBGnewfig}b) at any small coupling. 
Consequently, the Wannier center winding disappears, the band topology is trivialized, and one can construct localized Wannier functions. 
Very recently it was realized~\cite{ourpaper2} that if one further imposes the particle-hole symmetry $P$ on TBG, the topology becomes stable and not fragile. The combined antiunitary operator $(P C_{2z}\TRS)^2=-1$ mimics the case of spinful time-reversal symmetry and protects Kramers doublets in the Wilson spectrum at $k_{M_1}=0$ and $k_{M_1}=\pi$ \cite{yu_equivalent_2011,Aris2014WL} (see Fig.~\ref{TBGnewfig}d). 
The flow of the Wannier centers in TBG thus shows winding, Figs.~\ref{TBGnewfig}c,d, and the valley-$K$ model Eq.~\eqref{eq:Ham} is \emph{always} strongly topological no matter how many bands are included. 
Mixing the other valley renders the topology trivial; at small angles, however, the two valleys do not mix \cite{bistritzer_moire_2011}.

\subsection{Lower bound of superconductivity in TBG from topology}
\label{sec:lower_bounds_TBG}

The winding of the Wilson loop reflecting the topology of the active bands in TBG \cite{song_all_2019} can also be characterized by an integer-valued invariant, Euler or Stiefel–Whitney class $e_2=1$~\cite{ahn_failure_2019}. Refs.~\cite{ahn_failure_2019, song_all_2019, ourpaper2} showed this implies the existence of $4l+2e_2 $ ($l$ integer) Dirac nodes at zero energy between the two active bands which cannot be energetically separated. Based on Sec.~\ref{Wfunctionoverlap}, the superfluid weight of topological bands should be nonzero; however, the inequality bound Eq.~\eqref{eq:ineq} cannot be used, as the Chern number 
is zero due to the $C_{2z}\TRS$ symmetry. 
Instead, the symmetry class of $(C_{2z}\TRS)^2=1$ provides a new lower bound for the supefluid weight in TBG \cite{xie2020}, using the universally true Eq.~\eqref{eq:Ds_flat}. 
The non-Abelian Berry connection and curvature of the two TBG flat bands can  be written as in Box 1, under a proper \emph{local} gauge choice on a patch in the BZ~\cite{song_all_2019,ahn_band_2018,zou2018, po_faithful_2019}.  
In \cite{xie2020} it was proved that the Wilson loop winding number \cite{song_all_2019} is equivalent to the Euler class $e_2$ in Ref.~\cite{ahn_band_2018} 
whose expression is given in Box [\ref{abelianQGT}].

Using the positive definiteness of the non-Abelian QGT $\mathfrak{G}_{ij}$ (choosing the vectors $c_x$ and $c_y$ in Box 1 properly), it was shown~\cite{xie2020} that the quantum metric is bounded by the off-diagonal part $f_{xy}$ of the non-Abelian Berry curvature $\mathcal{F}_{xy}$: ${\rm tr}\,g \geq 2|f_{xy}|$, which in turn means that the superfluid weight is bounded by the Euler characteristic:
\begin{equation}\label{eqn:c2tbound}
\begin{split}
&  \frac{1}{4\pi} \int_{\rm BZ} d^2k \,{\rm tr}\,g(\mathbf{k}) \geq \frac{1}{2\pi}\int_{\rm BZ'} d^2k\,|f_{xy}| \\ \nonumber &\geq\Bigg{|}\frac{1}{2\pi}\int_{\rm BZ'} d^2 k\,f_{xy}\Bigg{|} = |e_2|\,.
\end{split}
\end{equation}

\subsection{Quantum Geometry in the Correlated Insulator States of TBG} \label{correlatedI}

Remarkably, a generalization of the quantum metric also appears in the TBG correlated insulator states \cite{ourpaper5}. Being in the strong coupling limit, TBG interacting states of matter \cite{kang_strong_2019,bultinck_ground_2020,po_origin_2018,xie2020,julku2020,hu2019,kang_nonabelian_2020,soejima2020efficient,pixley2019,xie_HF_2020,liu2020theories,cea_band_2020,daliao2020correlation,abouelkomsan2020,repellin_FCI_2020,vafek2020hidden,fernandes_nematic_2020} depend chiefly on the eigenstates (but not the energies) of the flat bands.
The insulating states in TBG at integer filling are amongst the best studied experimentally and theoretically, and the least theoretically controversial as extensive theoretical efforts have been aimed at their explanation \cite{kang_strong_2019,bultinck_ground_2020,po_origin_2018,kang_nonabelian_2020,soejima2020efficient,pixley2019,xie_HF_2020,liu2020theories,cea_band_2020,daliao2020correlation,abouelkomsan2020,repellin_FCI_2020,vafek2020hidden,fernandes_nematic_2020}.  

The strong coupling TBG Coulomb Hamiltonians of Kang and Vafek \cite{kang_strong_2019} and others \cite{zou2018,bultinck2020, ourpaper3, ourpaper4} (similarly to the negative $U$ projected Hubbard models in the superfluid weight), allow for exact expressions of energy and eigenstate of the charge $\pm 1$ excitation (relevant for transport gaps), of neutral excitation, and of charge $\pm 2$ excitation (relevant for possible Cooper pair binding energy~\cite{ourpaper4, vafek2020hidden}). Ref.~\cite{ourpaper4,vafek2020hidden} found that the charge $1$ excitation dispersion is governed by a generalized QGT of the projected bands, convoluted with the Coulomb interaction.
Hence the QGT influences the charge excitation dispersion of the insulating states of TBG, similar to how it governs the Cooper pair dispersion for the attractive Hubbard model. Hence it seems that whether superconductivity develops out of the topological bare flat bands, or if it develops upon doping the dispersive excitation of the insulating state, the QGT plays a crucial role in both scenarions: it either gives nonzero superfluid weight or it created large quasiparticle dispersion, respectively.

\section{Multilayers and flat bands in ultracold gases}

\begin{figure*} [ht!]
	\includegraphics[scale=0.5]{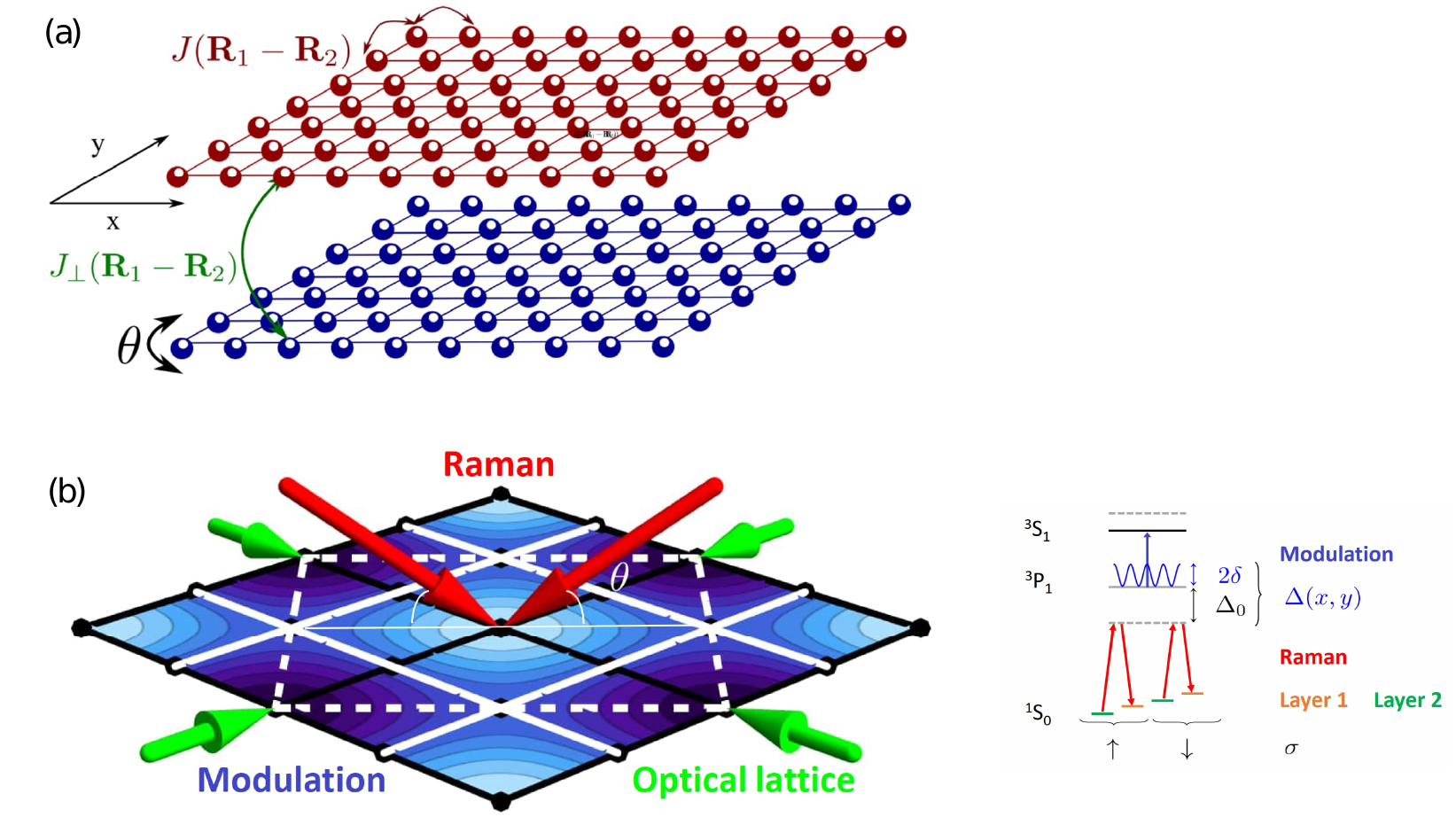}
\caption{\label{fig:ultracold_gas_TBG}\textbf{Ultracold gas analogues of twisted bilayer graphene (TBG).}
	 \textbf{(a)}  Two square optical lattices (red and blue) with an angle $\theta$ between each other form the analogue of the two layers in TBG. The two optical lattices are displaced in the $z$ direction for clarity, in reality they would be spatially overlapping and trap separately two long-lived atomic states, which play the role of the layer degree of freedom. There is no interference pattern between the two optical lattices since the corresponding light fields have orthogonal polarizations. The intralayer hopping $J$ is tuned by adjusting the depth of the two optical lattices, while a tunable interlayer hopping $J_{\perp}$ can be generated using either a direct transition, using microwave radiation, or an indirect Raman transition between the atomic states. Schematic of the proposal of Ref.~\onlinecite{Gonzalez-Tudela2019}.
	 \textbf{(b)} A single state-insensitive optical lattice (green) with spatially dependent  Raman coupling. This is obtained with a spatially varying detuning $\Delta(x,y)$ of the Raman beams induced by a \qql modulation laser\qqr (blue).  This has the effect of modulating the interlayer hopping $J_\perp$ in a way similar to the spatial twisting of the optical lattices in panel a. Panel a adapted with permission from Ref.~\onlinecite{Gonzalez-Tudela2019}. Panel b adapted with permission from Ref.~\cite{Salamon2020}}
\end{figure*}

Atoms  trapped by magnetic fields and light-induced potentials can be cooled to extremely low temperature, and by means of laser standing waves it is possible to force the atoms to move on the discrete lattice formed by the local minima of the optical potential, thereby mimicking the lattice of atomic orbitals in solids~\cite{Bloch2008,Giorgini2008,Torma2015,Lewenstein2012,Cooper2019}. These so-called optical lattices are  just one example of the high degree of tunability of ultracold gases, which makes them  an excellent quantum simulation platform to tackle currently intractable problems in quantum many-body physics and condensed matter physics. After the realization  of Bose-Einstein condensates and quantum degenerate Fermi gases,  experiments with ultracold gas have become increasingly sophisticated. The rich and versatile ultracold gas experimental toolbox has allowed to realize, among other things, artificial graphene lattices~\cite{Soltan-Panahi2011,Tarruell2012}, geometries displaying flat bands like kagome~\cite{Jo2012} and Lieb~\cite{Taie2015} lattices, bilayers~\cite{Gall2021}, quasicrystals~\cite{Sbroscia2020}, and synthetic dimensions~\cite{Mancini2015,Stuhl2015,Ozawa2019}. Moir\'e patterns have been predicted to occur also in vortex lattices in Bose-Einstein condensates~\cite{ORiordan2016}.

The phase diagram of TBG depending on the twist angle and filling is currently a hotly debated subject, and proposals have been put forward recently for realizing an ultracold gas analogue of TBG~\cite{Gonzalez-Tudela2019,Salamon2020,Salamon2020a,Luo2021} and investigating this problem in a well-controlled setting. The central idea common to all proposals is to use the internal hyperfine states of the atoms to encode the layer degree of freedom of TBG as a synthetic dimension. In practice this means that the optical potential depends on the hyperfine state. In this way it is possible to realize optical lattices for different atomic states that are twisted with respect to each other (Fig.~\ref{fig:ultracold_gas_TBG}a). There are different methods to realize a state-dependent optical potential, but they all rely on a clever arrangement of the laser field polarizations and the level structure of the atoms. The interlayer tunneling is engineered either by coupling the two states directly with microwave radiation or by a two-photon Raman transition. In alternative, it is possible to obtain the same result by using a single state-insensitive optical lattice together with a spatially modulated Raman coupling~\cite{Salamon2020} (Fig.~\ref{fig:ultracold_gas_TBG}b). In the latter case the period of the \MR{} pattern is precisely the period of the modulation of the Raman detuning and one can avoid  to physically twist the optical lattices.

The first strategy of tilted state-dependent optical lattices has been used in a recent experimental implementation with Rubidium-87~\cite{Meng2021}. Whereas $^{87}$Rb is bosonic, a generalization of the same scheme for fermionic atoms could allow to realize a  faithful model
of TBG in the near future. These recent results demonstrate that twisted bilayers are feasible within the ultracold gas toolbox, however to probe the effect of quantum geometry on superfluidity it is also necessary to measure observables such as the superfluid weight, in particular its dependence on the interaction strength. A whole set of new ideas and methods are required for measuring such observables in ultracold gases~\cite{Carusotto2011}. A lot of work has already been done in this direction: in the case of the unitary Fermi gas the superfluid fraction has been measured by directly imaging the density variations associated to the propagation of a wave of second sound~\cite{Sidorenkov2013}, a mode characterized by out of phase oscillations of the normal and superfluid components. 

Some theoretical ideas for measuring the superfluid density are a direct analog of the classical experiment with helium-II by Andronikashvili: a finite angular velocity is imparted to the ultracold gas, either by rotating the optical trap or by using a light-induced vector potential, and then the moment of inertia, directly related to the superfluid fraction, is obtained from the density redistribution~\cite{Ho2010} or spectroscopically from the hyperfine level population~\cite{John2011,Edge2011}. Other approaches based on artificial gauge fields can possibly allow to measure the superfluid density at a local level and out of equilibrium~\cite{Carusotto2011}. The ability to bring the ultracold gas out of equilibrium by suddenly changing a control parameter (a so-called quench) can also be useful to measure the superfluid weight in lattice systems~\cite{Peotta2014,Rossini2014}. Another route to probe the transport properties in a rather direct fashion is to realize atomtronic two-terminal setups~\cite{Krinner2017,Krinner2015}, such as the one that has allowed to observe superfluid flow in the presence of disorder~\cite{Krinner2013}. In a two-terminal setup superfluid transport can be probed also through the Josephson effect~\cite{Pyykkonen2021}.

\section{Outlook} 

The fingerprint of quantum geometric flat band superconductivity is the linear dependence on the interaction of the following quantities: Cooper pairing temperature, the pairing gap, superfluid weight, and the BKT superfluid transition temperature -- the linear dependence holds when the interaction is smaller than or similar to the hopping energies (see Section~\ref{sec:BKT_flat _band}). Experiments where (at least some of) these quantities are measured while varying the interaction would reveal – if approximately linear dependence is seen -- the geometric contribution of superfluidity for the first time. The normal state above the critical temperature is also likely to reveal interesting preformed pair, pseudogap or insulator behavior characteristic for a flat band~\cite{Tovmasyan2018,jiang_charge_2019,WangLevin2020,Peri2021,Huhtinen2021}.  

As platforms, the 2D moir\'e materials~\cite{MacDonald2019,Andrei2020,Balents2020,Kennes2021,Andrei2021} and ultracold gases~\cite{Bloch2008,Giorgini2008,Torma2015,Lewenstein2012} are complementary. Although remarkable demonstration on tuning the electronic interactions via electric field exists in twisted graphene~\cite{Park2021}, in general the control of interaction is easier and more precise in ultracold gases. There, the Feshbach resonance technique~\cite{Chin2010} allows realizing negative or positive interaction strengths of any magnitude, without affecting any other system parameter. The microscopic interaction is well approximated by a contact interaction, allowing to test Hubbard-type theories precisely. Another advantage of ultracold gas setups is that the ratio between inter- and intra-layer tunneling amplitudes can be adjusted more freely than in \MR{} materials. This opens up the
possibility of observing quasi-flat bands for larger rotation angles (smaller \MR{} unit cells), for instance. The great advantage of the 2D materials is that the temperatures required for superconductivity can actually be reached. While continuum ultracold gases routinely show superfluidity, bringing \textit{fermions in lattices} to the superfluid temperatures is a challenge, although advances in realizing antiferromagnetic states~\cite{Mazurenko2017} suggest it should be reached soon. While waiting, ultracold gases suit for exploring the normal state, and bosonic superfluidity. Ultracold gases are clean systems where impurities can be added in a controllable way, so they are an excellent platform for understanding the effect of disorder. The quantum gas microscopes~\cite{Torma2015} provide direct information about the fluctuations and quantum correlations of the ground state. For TBG, revealing quantum geometry effects via noise measurements has been proposed~\cite{Dolgirev2021}. The charge quasiparticle dispersion which is related to QGT (Section \ref{correlatedI}) can be revealed by STM. Concerning transport experiments, 2D materials are clearly the advantageous platform, although opportunities exist also in ultracold gases~\cite{Krinner2015,Pyykkonen2021}. For TBG, experimental work exploring the role of quantum geometry in superconductivity has already started~\cite{Tian2021}.  

Superconductivity in TBG is obviously only the beginning. Another promising \MR{} material is $\mathrm{WTe}_2$, where a topological exciton insulator has been observed in single layer \cite{jia2020evidence} but where metallic behavior is obtained in \MR{} bilayers \cite{wang2021onedimensional}; the appearance of superconductivity at some twist angles in $\mathrm{WTe}_2$ would be interesting due to its topological bands. Transition to the superconducting state has been observed already in magic angle twisted trilayer graphene (MATTG)~\cite{Park2021}.
Remarkably, the transition temperature in MATTG was more than 2 K, larger than in MATBG. Would stacking more layers eventually lead to room temperature superconductivity, in the spirit of the early proposal~\cite{kopnin2011}? Certainly, materials which have been thought to be non-superconducting in the past -- such as untwisted graphene, become superconducting when three layers are stacked together~\cite{zhou2021superconductivity}. Defining the highest possible critical temperature is an important research topic; recent works suggest existence~\cite{Hazra2019,Verma2021} or absence~\cite{Hofmann2021} of an upper bound.  Due to the fundamental lower bound of superfluidity Eqs.~\eqref{eq:Ds_flat}, \eqref{eq:ineq}, one can expect higher critical temperatures for large quantum metric, or Chern number. However, these results assume the isolated flat band limit; the quantum metric typically diverges when bringing other bands close to the flat band. While the divergence does not show up in the superfluid weight (due to the  contribution of other bands), it still leads in general to the enhancement of the latter~\cite{julku2016}. This hints to a possible sweet spot where quantum geometry effects are maximized by bringing other bands close to the flat band, while keeping them far enough so that most of the pairing takes place in the flat band where interactions dominate. 

The quantum metric is a fundamental quantity describing distances between the eigenstates of a system, and hence appears in many observables of interacting systems. For instance, light-matter interactions~\cite{Topp2021,Chaudhary2021} and exciton condensates~\cite{hu2020excitoncondensate} in \MR{}~materials reflect the underlying quantum geometry as well. Recently, quantum geometry was predicted to stabilize Bose-Einstein condensates in flat bands~\cite{Julku2021}, relevant for bosonic condensates in ultracold gas and polariton systems, or even for 2D \MR{} materials at the bosonic end of the BCS-BEC crossover~\cite{Park2021}. \\ \\

\section*{Key points}
\begin{itemize}
\item
Bands of (quasi-)flat dispersion dramatically enhance Cooper pairing as their (nearly-)vanishing kinetic energy allows interaction effects to dominate. 
\item
Superfluidity and stable supercurrents are possible in a flat band if the band has non-trivial quantum geometry: the related overlap of the Wannier functions facilitates movement of interacting particles even when non-interacting particles would be localized.
\item
Twisted bilayer graphene (TBG) exhibits nearly flat bands at its Fermi energy for small twist angles. Theory work suggests that quantum geometry is essential for the experimentally observed TBG superconductivity, and that a topological invariant called Euler class provides a lower bound for its superfluid weight.
\item
Ultracold gases offer another promising platform for highly controllable studies of superfluidity in \MR{} geometries. The first experiments are on the way.
\item
A flat band dispersion together with a quantum geometry that guarantees superfluidity are powerful guidelines for the search of superconductivity at elevated temperatures.
\end{itemize}

\acknowledgements

S.P. and P.T. acknowledge support by the Academy of Finland under project numbers 330384, 336369, 303351, and 327293. B.A.B. acknowledges support from the Office of Naval Research grant No. N00014-20-1-2303 and from the European Research Council (ERC) under the European Union's Horizon 2020 research and innovation programme (grant agreement n° 101020833). \\

\section*{Author contributions}
All authors have contributed to the writing of the manuscript.

\section*{Competing interests}
The authors declare no competing interests.

\bibliography{refs,bib_tbg,bec_paper2019,biblio_Nat_Phys_Rev_Sebastiano,TBLG.bib,Nature_review_geometric_superfluidity}

\end{document}